\documentclass[openacc]{rstransa}

\usepackage{graphicx}
\usepackage{indentfirst}
\usepackage{psfrag}
\usepackage{epsfig}
\usepackage{amsmath}
\usepackage{amssymb}
\usepackage{bm}
\usepackage{mathrsfs}
\newcommand{\be}{\begin{eqnarray}}
\newcommand{\ee}{\end{eqnarray}}

\newcommand{\hel}{\mathscr{ H}}

\newcommand{\flow} {\mathscr{ F}}
\newcommand{\perd}{\mathscr{ P}}

\begin{document}

\title{Chiral optical fields: A  unified formulation of helicity scattered from particles and dichroism enhancement}

\author{Manuel Nieto-Vesperinas}

\address{Instituto de Ciencia de Materiales de Madrid, Consejo Superior de
Investigaciones Cient\'{i}ficas,\\
 Campus de Cantoblanco, Madrid 28049, Spain.}

\subject{optics}


\corres{Manuel Nieto-Vesperinas\\
\email{mnieto@icmm.csic.es}}

\begin{abstract}
 We establish a general unified formulation which, using the optical theorem of electromagnetic helicity, shows that  dichorism is a phenomenon arising in any  scattering -or diffraction-  process,  elastic or not, of chiral electromagnetic fields by objects either chiral or achiral. It is shown how this approach paves the way to overcoming well-known limitations of  standard circular dichroism, like its weak signal or the difficulties of using it with magnetodielectric particles.

Based on the angular spectrum representation of optical fields with only right circular or left circular plane waves, we introduce beams with transverse elliptic polarization and posessing a longitudinal component. Then our  formulation  for general optical fields  shows how to enhance  the extinction rate of incident  helicity, (and therefore the dichroism signal), versus  that of energy of the light scattered or emitted by a particle, or viceversa.
\end{abstract}




\maketitle

\section{Introduction}

Chiral fields are acquiring increasing attention due to their potential   as probes of matter at the nanoscale \cite{circpolemission,allenlibro,babiker,tang1,tang2,cameron1,circpollibro1,allenlibro1,barnetspectr},
 of which life molecules are of paramount importance, or  as  high information capacity  signals in communication channels  \cite{allenlibro1,andrews1,andrews2,boyd1, zeilinger1}   with  control and transfer of angular momentum, which includes recently developed structured materials and  metasurfaces. The   conservation of the electromagnetic  {\it helicity} of wavefields  (or, equivalentely, {\it  chirality} when they are quasi-monochromatic; we shall indistinctly use both terms for such fields) \cite{lipkin,tang1,cameron1,bliokh1} was recently shown \cite{nieto1} to lead to a new   {\it optical theorem} which characterizes the excitation and emission of field  helicity -or chirality-  by bodies, and that we believe  should play a growing relevant role in coming years with the progress of research on applications of twisted light.  

In this context,  we pointed out \cite{nieto1} that  {\it circular dichroism} (CD) \cite{schellman,craig,barron1},  i.e. the difference in absorption -or emission- of  energy  by molecular objects according to the handness of circularly polarized light (CPL),  is a particular case of this optical theorem for scatterers,  and hence it does not need to resort to quantum mechanics as usually done in its standard formulation. Thus this phenomenon is just a consequence of the conservation of helicity of electromagnetic fields on scattering.

Different studies have discussed what kind of structures are necessary to produce chiral fields and  whether CD requires those objects being chiral. However, some works have recently shown that this effect can be obtained with achiral objects \cite{zambra}. Moreover, separating the existence  of chirality from dichroism effects may be a problem in some observations \cite{norris}.  Nonetheless no general and unified framework, not limited to particular structures,  has been yet established.

 In this paper we show  that dichorism is not only an effect  due to absorption and e.g. fluorescent re-emission by molecules; but it constitutes a property of any  scattering interaction,  elastic or not, of electromagnetic twisted fields. Thus based on the aforementioned optical theorem for the helicity, we generalize the concept of dichroism and demonstrate how  it appears not only with CPL waves, but  also with arbitrary chiral optical fields. This allows the  design of an illumination that enhances the information content of the scattered signal, overcoming well-known limitations of  standard CD detection, like its weak signal or its difficulties with magnetic objects \cite{choi}.

For comprehensiveness we next present a summary of concepts associated to the helicity  and its optical theorem. Then we show how general optical fields, expressed by its angular spectrum of plane waves, may be represented as a superposition of CPL components  of right handed (RCP) and/or left-handed (LCP)  polarization. This explicitely formulates in a quantitative manner previous  descriptions of helicity of general wavefields; and permits us to introduce a class of  elliptically polarized  hypergeometric beams, as well as their Hermite and Laguerre  derivations, which naturally appear when such representation is applied to a Gaussian angular spectrum.

We then establish how the  helicity optical theorem, applied to arbitrary fields and to chiral optical beams in particular, leads to a unified generalization of the theory of dichroism. A first consequence of which is {\it to  put forward the way of enhancing} either {\it the extinction of helicity} and hence  {\it the dichroic signal}, or {\it the extinction of intensity}. Such configurations and  detections are amenable  to future experiments.

\subsection{The excitation of helicity}
Quasimonochromatic fields have a time-harmonic dependence, i.e. their electric and magnetic vectors ${\bf \cal E}$ and ${\bf \cal B}$ are described in terms of their complex representations  ${\bf E}$ and  ${\bf B}$ as:  ${\bf \cal E}({\bf r},t)=\Re [{\bf E}({\bf r}) \exp(-i\omega t)]$ and  ${\bf \cal B}({\bf r},t)=\Re [{\bf B}({\bf r}) \exp(-i \omega t)]$, respectively.  $\Re$ denoting real part. Then the two fundamental quantities we deal with in this work are the  {\it helicity density}, $\mathscr{ H}$, and the {\it density of flow of helicity},  $\mathscr{ F}$, which in a non-absorbing dielectric medium of    permittivity $\epsilon$,  permeability $\mu$ and refractive index $n=\sqrt{\epsilon \mu}$  are \cite{cameron1,nieto1}:
$\hel= <\hel>=\frac{1}{2k}\sqrt{\frac{\epsilon}{\mu}} \Im ({\bf E}\cdot {\bf B}^*)$ and $\flow= <\flow> =  \frac{c}{4n k } \Im ( \epsilon {\bf E}^* \times {\bf  E}  + \frac{1}{\mu}{\bf  B}^*\times \bf B) $.  $\,$ $<\cdot>$ denoting time-average, $\Im$  meaning imaginary part and $k=n \omega/c$. It must be recalled that for these  time-harmonic fields $\flow$ coincides with the {\it spin angular momentum density} and is \cite{tang1,cameron1,bliokh1, nieto1}  $k^2$ times the {\it flow of chirality}. On the other hand, $\hel$ is $k^2$ times the {\it chirality}.  Also, they fulfill the continuity equation  
 \cite{tang1,cameron1,bliokh1,nieto1}: $\dot{\hel}+ \nabla \cdot \flow = - \perd$ .  Where the helicity dissipation on  interaction of the fields with matter is represented by $\perd$.

Let a quasimonochromatic field, whose space-dependent complex representation is denoted as ${\bf E}_i, {\bf B}_i$,  illuminates a particle which we  consider magnetodielectric and bi-isotropic  \cite{kong,nieto3}, dipolar in the wide sense i.e.  if for instance it is a sphere, its  magnetodielectric response is characterized by its electric,  magnetic, and magnetoelectric polarizabilities: $\alpha_{e}$,  $\alpha_{m}$,  $\alpha_{em}$,  $\alpha_{me}$, given by the first order Mie coefficients as: $
\alpha_{e}=i\frac{3}{2k^{3}}a_{1}$,  $\alpha_{m}=i\frac{3}{2k^{3}}b_{1}$,  $\alpha_{em}=i\frac{3}{2k^{3}}c_{1}$,  $\alpha_{me}=i\frac{3}{2k^{3}}d_{1}=-\alpha_{em}$. $\,$  $a_{1}$, $b_{1}$ and  $c_{1}=-d_{1}$ standing for the electric, magnetic, and magnetoelectric first Mie coefficients, respectively \cite{nieto3,bohren}. The conditiion   $\alpha_{em}=-\alpha_{me}$ expressing that the object is chiral. $\,$ We remark that by {\em particle} we shall understand small objects such as e.g. atoms, molecules,  material
macroscopic particles, or quantum dots.

The electric and magnetic dipole moments, $ {\bf p}$ and  ${\bf m}$, induced in the particle by this incident field are:
\be
{\bf p}=\alpha_{e} {\bf E}_i+\alpha_{em}{\bf  B}_i, \,\,\,\,\,
{\bf m}=\alpha_{me}{\bf E}_i+\alpha_{m}{\bf B}_i. \label{consti}
\ee
At any point outside this scattering object, the total field is written as:  ${\bf E}({\bf r})={\bf E}_i({\bf r})+{\bf E}_s({\bf r})$,   ${\bf B}({\bf r})=+ {\bf B}_s({\bf r})$. The subindex $s$ denoting the scattered, or radiated,  field.

 The  {\it optical theorem} that rules the {\it conservation of helicity} described by the above mentioned equation:  $\dot{\hel}+ \nabla \cdot \flow = - \perd$, $\,$   is \cite{nieto1}
\be
-{\cal W}_{\hel}^{a}=\frac{8\pi c k^3}{3 \epsilon} \Im[{\bf p} \cdot {\bf m}^{*}]- 
\frac{2\pi c}{\mu} \Re \{ -\frac{1}{\epsilon} {\bf p}  \cdot {\bf B}_{i}^{*} +\mu  {\bf m}\cdot  {\bf E}_{i}^{*}  \}. \,\,\,\,\label{todip4}
\ee
In the left side of  (\ref{todip4}) ${\cal W}_{\hel}^{a}$ is the rate of {\it  dissipation} by the particle of the incident field helicity. It comes from the integration of $\perd$ in a volume that contains this body. On the other hand, from Gauss' divergence theorem the terms  in the right side  of  (\ref{todip4}) arise from  the flow of $\flow$ across a surface that contains the particle \cite{nieto1}. The first of these terms represents the {\it  total helicity scattered or radiated} by the object, whereas the second one constitutes the {\it extinction of helicity} of the incident wave on scattering. This latter extinction term: $-(2\pi c/\mu) \Re \{ -\frac{1}{\epsilon} {\bf p}  \cdot {\bf B}_{i}^{*} +\mu  {\bf m}\cdot  {\bf E}_{i}^{*}  \}$  should be used for determining both dissipated and radiated, or scattered, helicity by a dipolar particle in an arbitrary, homogeneous or inhomogeneous, embedding medium.  To emphasize this interpretation  we recall its analogy with the well-known  optical theorem for energies \cite{born}
\be
 -{\cal W}^{a}=\frac{c k^4}{3 n} [\epsilon^{-1} |{\bf p}|^2 + \mu |{\bf m}|^{2}] 
-\frac{\omega}{2} \Im[ {\bf p}  \cdot {\bf E}_{i}^{*}   +{\bf m}\cdot  {\bf B}_{i}^{*}] . \label{top}
\ee
${\cal W}^{a}$ being the rate of energy absorption from the illuminating wave. In  the right side of (\ref{top}) the first term constitutes the total energy scattered by the dipolar object, whereas the second one represents the energy extinguished  from the illuminting field, or  rate of energy excitation in the scattering object. 

Henceforth we remark the analogous role played by the right side terms in both optical theorems (\ref{todip4}) and (\ref{top}). As is well-known  $(\omega/2)\Im [{\bf p} \cdot {\bf E}^*+{\bf m} \cdot {\bf B}^*]$ has been extensively employed for characterizing dipole optical interactions \cite{novotny,  klimov, dionne}. We thus expect  that   progress on  research of radiation-matter interactions with chiral fields will give rise to a growing use of the helicity extinction in  Eq. (\ref{todip4}): $(2\pi c/\mu) \Re \{ -\epsilon^{-1} {\bf p}  \cdot {\bf B}_{i}^{*} +\mu  {\bf m}\cdot  {\bf E}_{i}^{*}  \}$. Based on this reasoning we find it natural to introduce an {\it  enhancement factor} $F_{\hel}$ for the emission of helicity in analogy with the   Purcell factor for a radiating electric and/or magnetic dipole: $F=1+(3/2k^3) [\Im \{{\bf p} \cdot {\bf E}^* +{\bf m}\cdot {\bf B}^*\}]/ [{\epsilon^{-1}|{\bf p}|^2  +\mu |{\bf m}|^2}] $, viz. : 
\be
F_{\hel}=1+\frac{3\epsilon}{4\mu k^3} \frac{\Re \{ -\frac{1}{\epsilon} {\bf p}  \cdot {\bf B}_{i}^{*} +\mu  {\bf m}\cdot  {\bf E}_{i}^{*}  \}}{\Im[{\bf p} \cdot {\bf m}^{*}\}}. \label{FH} 
\ee
In this connection, and  analogously to the complex Poynting  vector theorem of energy conservation, (see Sec 6.10 of  \cite{jackson} and also \cite{norris}),  the integration of the above mentioned continuity equation   for a lossy particle of volume $V$ with constitutive parameters $\epsilon=\epsilon_R + i \epsilon_I$ and $\mu=\mu_R+i\mu_I$, in absence of induced currents,  yields (\ref{todip4}) with: ${\cal W}_{\hel}^{a}=\frac{c^2}{2 n^2}\int_V dv (\epsilon_R \mu_I + \epsilon_I \mu_R) \Im\{\frac{{\bf E}\cdot  {\bf B^*}}{\mu^*}\}; $ which  links fields  in, or close to, the object  with  tose  in any other region of space; in particular in the far-zone.

\section{The angular spectrum of  circularly polarized plane wave components}
We address the wide variety of fields propagating in a half-space $z>0$, or $z<0$, free from sources, represented by an angular spectrum of plane waves \cite{mandel, nietolib}. This includes optical fieds. Such representation of either  incident and scattered  fields, with subindex $i$ and $s$ respectively, is: 
\be
{\bf E}_{i,s}({\bf r})=\int_{\cal D} {\bf e}_{i,s}({\bf s_{\perp}}) e^{ik({\bf s}\cdot {\bf r})} d\Omega, \,\,\,\, \,\,\,\,
{\bf B}_{i,s}({\bf r})=\int_{\cal D} {\bf b}_{i,s}({\bf s_{\perp}}) e^{ik({\bf s}\cdot {\bf r})} d\Omega.  \label{ang1}
\ee
The integration being done on the  contour $\cal D$ that contains both propagating and evanescent waves \cite{mandel,nietolib}.  ${\bf s}=(\bf s_{\perp}, s_{z}) $ is the unit wavevector of  the plane wave component of amplitude ${\bf e}_{i,s}({\bf s}_{\perp})$ and  ${\bf b}_{i,s}({\bf s}_{\perp})$ , where ${\bf s}_{\perp}=(s_x,s_y,0)$ and   $s_{z}=\pm \sqrt{ 1-|{\bf s}_{\perp}|^2}$  if  $|{\bf s}_{\perp}|^2 \leq 1$, (propagating components); and   $s_{z}= \pm i \sqrt{ |{\bf s}_{\perp}|^2-1}$ if   $|{\bf s}_{\perp}|^2 > 1$, (evanescent  components). $d\Omega=\sin \alpha d\alpha d \beta$. $s_{x}=\sin\alpha\cos\beta$, $s_{y}=\sin\alpha\sin\beta$, $s_{z}=\cos\alpha$. $0\leq \beta \leq 2\pi$,  $0\leq \alpha \leq \pi/2$ for propagating components and $ \alpha=\pi/2 - i \delta$, $0<\delta \leq \infty$ for evanescent components. The $+$ or $-$ sign in $s_z$ applies according to wether propagation is in $z>0$ or $z<0$, respectively. We shall assume the first case. For  $z<0$ the results are similar.
\begin{figure}[htbp]
\centerline{\includegraphics[width=0.6\columnwidth]{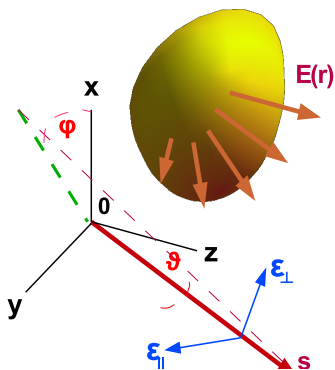}}
\caption{(Color online). A field ${\bf E({\bf r})}$, with wavefront shown by the brown-yellow surface, propagates into the half-space $z>0$ along arbitrary directions, (light-brown arrows).  In the  $0XYZ$ framework the propagation vector along ${\bf s}$ of each plane wave component of ${\bf E({\bf r})}$ has polar and azimuthal angles $\theta$ and $\phi$. The polarization of each of these plane waves is characterized by  the orthonormal system $\{{\bm \epsilon}_{\perp},{\bm \epsilon}_{\parallel}, {\bf s}\}$. The unit vector ${\bm \epsilon}_{\parallel}$ is in the polar plane containing both ${\bf s}$ and its projection (green broken line) on $OXY$, and points in the rotation sense   of $\theta$. On the other hand, ${\bm \epsilon}_{\perp}$ is normal to this plane and points against the sense of rotation of $\phi$. }
\end{figure}
In general all plane wave components are elliptically polarized. For the incident and scattered fields one has:
 ${\bf b}_{i,s}({\bf s}_{\perp}) =n {\bf s} \times  {\bf e}_{i,s}({\bf s}_{\perp}) $, ${\bf e}_{i,s}({\bf s}_{\perp})  \cdot {\bf s}= {\bf b}_{i,s}({\bf s}_{\perp})  \cdot {\bf s}=0$.  The complex amplitudes of the scattered, or radiated,  field angular spectrum being:
\begin{equation}
{\bf e}_s({\bf s}_{\perp})=k^2  [\epsilon^{-1}({\bf s} \times {\bf
p})\times {\bf s}- \sqrt{\frac{\mu}{\epsilon}}({\bf s}\times
{\bf m})]; \,\,\,\,\,\,\,\,\,
{\bf b}_s({\bf s}_{\perp})=k^2 [\mu({\bf s} \times {\bf m})\times
{\bf s}+\sqrt{\frac{\mu}{\epsilon}} ({\bf s}\times {\bf p})]. \label{dipm}
\end{equation}
 For each  plane wave component  with wavevector  $k{\bf s}$ of either the incident or the scattered field (\ref{ang1}), we consider an orthonormal  set of unit vectors  (cf. Fig. 1)  \{$\hat{\bm \epsilon}_{\perp},\hat{\bm \epsilon}_{\parallel}, {\bf s}\}$   from which   we  define an helicity basis of rotating vectors:   ${\bm \epsilon}^{\pm}({\bf s})=(1/\sqrt{2})(\hat{\bm \epsilon}_{\perp}({\bf s}), \pm i \hat{\bm \epsilon}_{\parallel}({\bf s}))$,   \,\,\,\,\,\, ${\bm \epsilon}^{\pm\,*}({\bf s})\cdot {\bm \epsilon}^{\mp}({\bf s})=0$. Then each incident or scattered component complex amplitude is expressed as the sum of a left-handed (LCP, sign "+")  and a right-handed  (RCP, sign "-") circularly polarized plane wave  in its corresponding framework  \{$\hat{\bm \epsilon}_{\perp},\hat{\bm \epsilon}_{\parallel}, {\bf s}\}$   according to 
 \be
{\bf e}_{i,s}({\bf s}_{\perp})=e_{i,s}^{+}({\bf s}_{\perp}){\bm \epsilon}^{+}({\bf s})+ e_{i,s}^{-}({\bf s}_{\perp}){\bm \epsilon}^{-}({\bf s}). \,\,\,\,\,\,\,\,\,\,  \label{eheli} \,\,\,\,\,\,\,\,\,\,\,\,\,\,\,\,\,\,\, \\ 
{\bf  b}_{i,s}({\bf s}_{\perp})
=b_{i,s}^{+}({\bf s}_{\perp}){\bm \epsilon}^{+}({\bf s})+ b_{i,s}^{-}({\bf s}_{\perp}){\bm \epsilon}^{-}({\bf s})\,\,\,=-ni[e_{i,s}^{+}({\bf s}_{\perp}){\bm \epsilon}^{+}({\bf s})- e_{i,s}^{-}({\bf s}_{\perp}){\bm \epsilon}^{-}({\bf s})].  \,\,\,\,\,\,\,\,\,\,\,\,\,\,\,\,\,\,\,\,\,\,\,\,\,\label{bheli}
\ee
With equation  ${\bm \nabla}\cdot {\bf E}=0$ imposing according to (\ref{ang1})  that  ${\bm \epsilon}^{\pm}({\bf s})\cdot {\bf s}=0$. 
In this representation, the  helicity density of each incident or  scattered plane wave component  reads:    
\be
{\hel^{i,s}}({\bf s}_{\perp})=(\epsilon/k)\Im[e_{i,s \, x}^{*}({\bf s}_{\perp})e_{i,s \, y}({\bf s}_{\perp})]= (\epsilon/2k) S_{3}({\bf s}_{\perp}) =(\epsilon/2k)[|e_{i,s}^{+}({\bf s}_{\perp})|^2-|e_{i,s}^{-}({\bf s}_{\perp})|^2]. \,\,\,\,\,\,\,\,\,\,\,\,\,\,\,\label{helhel}
\ee
 I.e. as the difference between the LCP and RCP intensities of  this angular component  of wavevector $k{\bf s}$. $\,$ $S_3({\bf s}_{\perp})$ is the 4th Stokes parameter \cite{born,marston1}. Also  $|e_{i,s}({\bf s})_{\perp}|^2= | e_{i,s\, x}({\bf s}_{\perp})|^2+|e_{i,s \, y}({\bf s}_{\perp})|^2=\frac{8\pi}{c}\sqrt{\frac{\mu}{\epsilon}} <S_{i,s}({\bf s}_{\perp})>=\frac{8\pi}{\epsilon} <w_{i,s}({\bf s}_{\perp})>$. $<S_{i,s}({\bf s}_{\perp})>$ and $<w_{i,s}({\bf s}_{\perp})>$ representing the   time-averaged Poynting vector magnitude  and electromagnetic  energy density, respectively: 

 $<w_{i,s}({\bf s}_{\perp})>=<w_{e \, i,s}({\bf s}_{\perp})>+<w_{m \, i,s}({\bf s}_{\perp})>$ .  $<w_{e \, i,s}({\bf s}_{\perp})>=(\epsilon/16 \pi)|{\bf e}_{i,s}({\bf s}_{\perp})|^{2}$, $<w_{m \, i,s}({\bf s}_{\perp})>=(1/16 \pi  \mu)|{\bf b}_{i,s}({\bf s}_{\perp})|^{2}$.  

 Therefore for the incident or the scattered field we have from  (\ref{ang1}), (\ref{eheli}) and (\ref{bheli}) the following  splitting into LCP and RCP waves
\be
{\bf E}_{i,s}({\bf r})={\bf E}_{i,s}^{+}({\bf r})+{\bf E}_{i,s}^{-}({\bf r});\,\,\,\,\,\,\,\,\,\,\,\,\,\,\
{\bf B}_{i,s}({\bf r})={\bf B}_{i,s}^{+}({\bf r})+{\bf B}_{i,s}^{-}({\bf r})
=-ni[{\bf E}_{i,s}^{+}({\bf r})-{\bf E}_{i,s}^{-}({\bf r})]. \label{elipt1} \\
{\bf E}_{i,s}^{\pm}({\bf r})=\int_{\cal D} e_{i,s}^{\pm}({\bf s}_{\perp}){\bm \epsilon}^{\pm}({\bf s}) e^{ik({\bf s}\cdot {\bf r})} d\Omega.  \,\,\,\,\,\,\,\,\,\,\,\,\,\,\,\,\,\,\,\,\,\,\,\,\,\,\,\,\,\ \label{ang masmen}
\ee
Assuming the particle chiral, $\alpha_{em}=-\alpha_{me}$, and introducing  Eqs.(\ref{ang masmen}) into (\ref{consti}) we write:
\be
{\bf p}({\bf r})={\bf p}_{+}({\bf r})+{\bf  p}_{-}({\bf r}); \,\,\,\,\,\,\,\,\,\,\,\,\,\,\
{\bf m}({\bf r})={\bf m}_{+}({\bf r})+{\bf m}_{-}({\bf r}).  \label{pm masmen}
\ee
With
\be
{\bf  p}_{\pm}({\bf r})=(\alpha_{e} \pm n i \alpha_{me}) {\bf E}_{i}^{\pm}({\bf r}). \,\,\,\,\,\,\,\,\,\,\,\,\,\,\ 
{\bf  m}_{\pm}({\bf r})=(\alpha_{me} \mp n i \alpha_{m}) {\bf E}_{i}^{\pm}({\bf r}). \label{constia}
\ee
And  substituting (\ref{ang1})  into (\ref{constia}) we see that ${\bf  p}_{\pm}({\bf r})$ and ${\bf  m}_{\pm}({\bf r})$ also admit an  angular spectrum representation like  (\ref{ang1}),  their respective angular spectra  being:
\be
\hat {\bf  p}_{\pm}({\bf s}_{\perp})=   (\alpha_{e} \pm n i \alpha_{me}) e_{i}^{\pm}({\bf s}_{\perp}){\bm \epsilon}^{\pm}({\bf s}); \,\,\,\,\,\,\,\,\,\,\,\,\,\,\,\,\,\,\,\,\,\,\,\,\,\,\,\,\,
 \hat{\bf  m}_{\pm}({\bf s}_{\perp})= (\alpha_{me} \mp n i \alpha_{m}) e_{i}^{\pm}({\bf s}_{\perp}){\bm \epsilon}^{\pm}({\bf s}). \label{pym_angul}
\ee
So that from (\ref{pym_angul}), (\ref{pm masmen}) and (\ref {dipm}) we obtain for the scattered field angular spectrum:
\be
e_{s}^{\pm}({\bf s}_{\perp})=k^2 [\frac{\alpha_{e} \pm n i \alpha_{me}}{\epsilon}\pm i \sqrt{\frac{\mu}{\epsilon}}(\alpha_{me} \mp n i \alpha_{m})] { e}_{i}^{\pm}({\bf s}_{\perp}). \,\,\,\,\,\, \nonumber \\
b_{s}^{\pm}({\bf s}_{\perp})=k^2 [\mp i \sqrt{\frac{\mu}{\epsilon}}(\alpha_{e} \pm n i \alpha_{me}) +\mu (\alpha_{me} \mp n i \alpha_{m})] { e}_{i}^{\pm}({\bf s}_{\perp}). \label{angs}
\ee
We obtain the helicity densities  ${\hel}_{i,s}$ for either the incident or scattered fields by introducing   (\ref{eheli}) and (\ref{bheli})  into  (\ref{ang1}), and inserting the result into the  definition introduced in Section 1(a): $\hel= <\hel>=\frac{1}{2k}\sqrt{\frac{\epsilon}{\mu}} \Im ({\bf E}\cdot {\bf B}^*)$. Then since after taking imaginary parts the cross-terms containing the  integrand factors $ni e_{i,s}^{+}({\bf s}_{\perp})e_{i,s}^{-\, *}({\bf s'}_{\perp}){\bm \epsilon}^{+}({\bf s})\cdot {\bm \epsilon}^{-\, *}({\bf s'})$ and $-n i e_{i,s}^{+\, *}({\bf s}_{\perp})e_{i,s}^{-}({\bf s'}_{\perp}){\bm \epsilon}^{+\, *}({\bf s})\cdot {\bm \epsilon}^{-}({\bf s'})$ cancel each other,   we finally get:
\be
{\hel}_{i,s}({\bf r})= \frac{\epsilon}{2k}[|{\bf E}_{i,s}^{+}({\bf r})|^2 -  |{\bf E}^{-}_{i,s}({\bf r})|^2]. \label{hel2} \label{ang2}
\ee
Equation (\ref{hel2})  introduced in the  optical theorem  for the helicity, (\ref{todip4}), accounts for all chirality effects due to the interaction of waves with dipolar particles, both in the propagating region (real $s_{i}^{z}$) of the angular spectrum,  as in the evanescent domain (imaginary $s_{i}^{z}$). The latter applies in particular for the interaction of plasmon polaritons with particles on metallic surfaces.

Expressions (\ref{hel2})  are  of particular importance in the far zone $kr\rightarrow\infty$, where \cite{mandel,nietolib}
 \be
{\bf E}_{i,s}^{\pm}(r \hat{\bm s})\approx  -(2\pi i/k) e_{i,s}^{\pm}(\hat{\bm s}_{\perp}){\bm \epsilon}^{\pm}(\hat{\bm s}) \exp(ikr)/r.  \label{asympt}
\ee 
$e_{s}^{\pm}(\hat{\bm s}_{\perp}){\bm \epsilon}^{\pm}(\hat{\bm s})$ playing the role of the  CPL complex amplitude for a radiated, or scattered, field, and   $\hat{\bm s}={\bf r}/r$  now bolonging to the domain of propagating components only. Dropping the subindices $i,s$ to  simplify notation, Eqs.(\ref{hel2}) and (\ref{asympt})  lead for either the incident or the scattered field to:
\be
{\hel}_{ff}(r \hat{\bm s})=\frac{2\pi^2 \epsilon}{k^3 r^2}[|{ e}^{+}(\hat{\bm s}_{\perp})|^2 -  |{e}^{-}(\hat{\bm s}_{\perp})|^2]. \label{hff}
\ee
And  their density of flow  of  helicity is: $\flow_{ff}(r \hat{\bm s})=\frac{c}{n}{\hel}_{ff}({\bf r})\hat{\bm s}$; which in agreement with the conservation of helicity, expresses on integration in a large sphere surrounding the scatterer that the outgoing helicity  flow of the field  across any plane $z=constant$, or closed surface, outside the scattering volume, which equals the  flow of helicity across any sphere at infinity, is  equal to  $c/n$ times the  total helicity enclosed by that sphere: $\int\int_{z=0}\flow ({\bf r}) \cdot \hat{\bm z}dx dy =\int_{r\rightarrow \infty}\flow_{ff}(r \hat{\bm s})\cdot {\bm r}\,r^2 d\Omega.$ Where now the solid angle $\Omega$ spans on  the whole sphere of real angles only.
Taking into account (\ref{hel2}), and in analogy with the flow of energy \cite{mandel,nietolib},  one sees that the evanescent components do not contribute to the flux of helicity across the plane $z=0$ in the half-space $z\geq 0$.
\subsection{A particular case: Incident elliptically polarized  plane wave}

The significance of the optical theorem (\ref{todip4}) for the helicity - or chirality - of wavefields is illustrated  considering one of the simplest and most employed configurations: one elliptically polarized incident plane wave  impinging on a dipolar particle with wavevector  $k{\bf s}_i$ along $OZ$. According to (\ref{eheli}) and   (\ref{bheli}) the fields are:

${\bf e}_{i}=( e_{i x},e_{i y},0)=e_{i}^{+}{\bm \epsilon}^{+}+ e_{i}^{-}{\bm \epsilon}^{-}.  \nonumber \\  
{\bf  b}_{i}= ( b_{i x},b_{i y},0)=n ( -e_{i y},e_{i x},0)=b_{i}^{+}{\bm \epsilon}^{+}+ b_{i}^{-}{\bm \epsilon}^{-} 
=-ni(e_{i}^{+}{\bm \epsilon}^{+}- e_{i}^{-}{\bm \epsilon}^{-}). $

So that the incident helicity density  reads:    
\be
{\hel^{i}}=(\epsilon/k)\Im[e_{i x}^{*}e_{i y}]= (\epsilon/2k) S_{3} 
=(\epsilon/2k)[|e_{i}^{+}|^2-|e_{i}^{-}|^2]. 
\ee
Also,  according to (\ref{pm masmen}) and (\ref{constia}): 

${\bf p}=p_{+}{\bm \epsilon}^{+}+ p_{-}{\bm \epsilon}^{-}, \,\,\,\, p_{\pm}=(\alpha_{e} \mp n i \alpha_{em}) {e}_{i}^{\pm}; \,\,\,\,
{\bf m}= m_{+}{\bm \epsilon}^{+}+ m_{-}{\bm \epsilon}^{-},  \,\,\,\,  m_{\pm}=(\alpha_{me} \mp n i \alpha_{m}) {e}_{i}^{\pm}.$

On introducing these dipole moments and fields into  the optical theorems of helicity (\ref{todip4})  and energy (\ref{top}), they  yield for the rate of   helicity and energy extinction:
\be
\Im\{(p_{+}+in m_{+}) {e}_{i}^{+ \,*}-(p_{-}-in m_{-}) {e}_{i}^{- \,*} \} 
=\frac{4k^3 n}{3\epsilon}\Im\{p_{+} m_{+}^{\,*} + p_{-} m_{-}^{\,*} \}+{\cal W}_{\hel}^{a} \label{topheli}
\ee
and
\be
\Im\{(p_{+}+in m_{+}) {e}_{i}^{+ \,*}+(p_{-}-in m_{-}) {e}_{i}^{- \,*} \} 
=\frac{2k^3 }{3\epsilon}\{|p_{+}|^2 +  |p_{-}|^2 +n^2 (|m_{+}|^2 + | m_{-}|^2 )\}+{\cal W}^{a}, \,\,\label{topene}
\ee
respectively. Eq. (\ref{topheli}) is identical to the CD  law, usually  mechanoquantically  formulating  molecular absorption and fluorescence effects   \cite{craig}.  However Eqs. (\ref{topheli})  and (\ref{topene}), obtained  from  classical electrodynamics,  include the rate of helicity and energy dissipation both by absorption and scattering (or diffraction),  and generalize the CD theory to any wide sense dipolar  "particle"  or structure.

In other words, {\it   the CD phenomenon  is not only characterized by  the  operation of taking  the  difference of energy absorption and emission   $\Im[ {\bf p}  \cdot {\bf E}_{i}^{*}   +{\bf m}\cdot  {\bf B}_{i}^{*}]$ by chiral molecules as they are  separately  illuminated by RCP  and  LCP waves; i.e as this absorbed energy  is  $ \Im\{(p_{+}+in m_{+}) {e}_{i}^{+ \,*}\}$ and $\Im\{(p_{-}-in m_{-}) {e}_{i}^{- \,*} \}$, respectively, as usually considered so far \cite{schellman, craig, barron1}. But CD is also, and fundamentally, one of the  physical manfestations of the conservation law of electromagnetic helicity -or chirality- and is  represented by the leftt-side  term of (\ref{topheli}),  $\Re[ -\frac{1}{\epsilon}{\bf p}  \cdot {\bf B}_{i}^{*}   +\mu {\bf m}\cdot  {\bf E}_{i}^{*}]= \Im\{(p_{+}+in m_{+}) {e}_{i}^{+ \,*}-(p_{-}-in m_{-}) {e}_{i}^{- \,*} \}$ of the helicity optical theorem  (\ref{todip4}); being involved in any  scattering and/or absorption process of LCP and RCP electromagnetic waves, thus characterizing the rate of extinction of helicity -or chirality-. In addition,  as shown by Eqs. (\ref{pym_angul}) and (\ref{topheli}), CD arises  not only due the chirality of the scattering object, represented by $\alpha_{me}$,  but also and primarily  by  the mere induction  of their electric and/or magnetic dipoles, whose responses are characterized by  $\alpha_{e}$ and $\alpha_{m}$,  respectively}.

Hence it is not surprising  that the ratio of the   extinction  of   incident field helicity (\ref{topheli})  and  energy (\ref{topene}) is identical to the well-known {\it  dissymmetry factor} of CD \cite{tang1,schellman,barron1}. Moreover, adding and substracting  (\ref{topheli})  and   (\ref{topene}) yield the energy excitation by extinction of the respective  LCP or RCP component of the incident elliptically polarized light according to the dipole handness $p_{\pm}$ and/or $m_{\pm}$:  
\be
\Im\{(p_{\pm}\pm in m_{\pm}) {e}_{i}^{\pm \,*} \}
=\frac{k^3 }{3\epsilon}\{|p_{+}\pm in m_{+}|^2 + |p_{-}\pm in m_{-}|^2\}+\frac{1}{2}({\cal W}^{a}\pm{\cal W}_{\hel}^{a}). \,\,\,\,\label{topeneheli0}
\ee
\subsection{The special case of an  incident circularly polarized plane wave}
Let the field incident on the particle be just one  CPL component, either LCP or RCP, then  $e_{i}^{\pm}=e {\bm \epsilon}^{\pm}$ and ${\bf p}=p_{\pm}{\bm \epsilon}^{\pm}$,   ${\bf m}= m_{\pm}{\bm \epsilon}^{\pm}$, and (\ref{topeneheli0}) lead to
\be
\Im\{(p_{\pm}\pm in m_{\pm}) {e}_{i}^{\pm \,*} \}
=\frac{k^3 }{3\epsilon}\{|p_{\pm}|^2 +n^2|m_{\pm}|^2 \pm 2n \Im\{p_{\pm}m _{\pm}^{*}\}\}+\frac{1}{2}({\cal W}^{a}\pm{\cal W}_{\hel}^{a}) . \,\,\,\,\label{topeneheli3}
\ee
and
\be
\frac{k^3 }{3\epsilon}|p_{\pm}\mp in m_{\pm}|^2 +\frac{1}{2}({\cal W}^{a}\mp{\cal W}_{\hel}^{a})=0. \,\,\,\,\label{topeneheli4}
\ee
From which we obtain
\be
2n \Im\{p_{\pm} m_{\pm}^{\,*}\}=\pm [|p_{\pm}|^2 +  n^2 |m_{\pm}|^2 +\frac{1}{2}({\cal W}^{a}\mp{\cal W}_{\hel}^{a}) ] . \,\,\,\,\label{topene6b}
\ee
Thus, apart from a constant factor,  for CPL incidence  the scattered helicity equals in modulus  the scattered energy plus the rates of dissipation of helicity and energy,  and has a sign that depends on the handness of the incident light. Of course   (\ref{topeneheli3}) - (\ref{topene6b}) are consistent, as they should, with Eqs.(\ref{topheli}) and (\ref{topene}), which for CPL become:
\be
\Im\{(p_{\pm}\pm in m_{\pm}) {e}_{i}^{\pm \,*}\} =\pm\frac{4k^3 n}{3\epsilon}\Im\{p_{\pm} m_{\pm}^{\,*}\} \pm {\cal W}_{\hel}^{a}
=\frac{2k^3 }{3\epsilon}\{|p_{\pm}|^2 +  n^2|m_{\pm}|^2 \}+{\cal W}^{a}. \,\,\,\,\label{topheli5}
\ee
A Comparison of  (\ref{topheli5})  with  (\ref{topheli}) and (\ref{topene}), shows that  the excitation of the particle by both the  LCP and  RCP components  of    an  elliptically polarized plane wave, is equivalent to performing two observations separately: one with an  LCP plane wave only, and another one with  only RCP, (each of which is ruled  by  (\ref{topheli5})  with the corresponding  sign),   and then substracting or adding the respective excitations  given by  the left sides of  (\ref{topheli5}). This operation reproduces the left side of (\ref{topheli}) and (\ref{topene}), respectively. In other words,  Eqs.(\ref{topheli}) and (\ref{topene}) show that  {\it the LCP and RCP components of an incident elliptically polarized plane wave do not interfere  and, hence,  interact independently of each other with the particle}. 

As regards Eq.  (\ref{topeneheli4}), since often in  molecular spectroscopy  $|m_{\pm}| <<| p_{\pm}|$,  the value of ${\cal W}^{a}$ and/or ${\cal W}_{\hel}^{a}$ contributes to that of $|p_{\pm}|$. Nonetheless Eq. (\ref{topeneheli4}) is also compatible with the electric and magnetic dipoles excited by CPL light, and the absorption rates, fulfilling:
\be
p_{\pm}=\pm  in m_{\pm} \Leftrightarrow  {\cal W}_{\hel}^{a} = \pm {\cal W}^{a}. \,\,\,\,\label{topeneheli5}
\ee
Hence {\it this is a sufficient condition for an electric-magnetic dipole to emit chiral light}. Particularly remarkable is this latter case is when  the dissipation rates of helicity and energy either cancel each other, or  the particle introduces no energy nor helicity losses, ${\cal W}^{a}={\cal W}_{\hel}^{a}=0$, so that all  energy and helicity extinguished from the incident field are re-radiated by elastic scattering. As seen from (\ref{topeneheli4}),  in that case: 
$2n \Im\{p_{\pm} m_{\pm}^{\,*}\}=\pm [|p_{\pm}|^2 +  n^2 (|m_{\pm}|^2 ] $, which states that then  the optical theorems for helicity, Eq. (\ref{topheli}), and  energy, Eq. (\ref{topene}), are equivalent, and  the scattered helicity is proportional to the scattered intensity  and has a sign that depends on the handness of the incident light, whereas the density of  helicity  flow (spin) is proportional to that of energy flow (Poynting vector).  Thus in such a  situation the optical theorems for helicity (\ref{todip4})  and  energy (\ref{top})  are equivalent, (see also \cite{cameron1,nieto1}).

Equation (\ref{topeneheli5}) also has some  important consequences:

\textbullet \,{\it The far-zone scattered field is circularly polarized}.    ${\bf b}^{\pm}({\bf s}_{\perp})=\mp n i{\bf e}^{\pm}({\bf s}_{\perp})$, [cf.  Eqs.    (\ref{dipm})]. This circular polarization  holds with respect to the Cartesian system of orthogonal  axes defined by the unit vectors: $({\bm \epsilon}_{\perp},{\bm \epsilon}_{\parallel}, {\bf s})$, (see Fig. 1). ${\bm \epsilon}_{\perp} $ and  ${\bm \epsilon}_{\parallel}$ being respectively  perpendicular and parallel to the polar plane (which now becomes the scattering plane)  delimited by ${\bf s}$ and its projection on $OXY$.  I.e.: ${\bf e}^{\pm}({\bf s}_{\perp})=({\bf e}({\bf s}_{\perp}) \cdot {\bm \epsilon}_{\perp})({\bm \epsilon}_{\perp}+ \pm i {\bm \epsilon}_{\parallel}+0{\bf s} )$ and ${\bf b}^{\pm}({\bf s}_{\perp})=( n {\bf e}({\bf s}_{\perp}) \cdot {\bm \epsilon}_{\perp})  (\mp i {\bm \epsilon}_{\perp}+ {\bm \epsilon}_{\parallel}+0{\bf s} )$.

From the above it should also be noticed  that {\it it is the handness of the dipole moments, and not necessarily  the chirality $\alpha_{me}$, the relevant characteristic for these CD  effects}. Besides, this CPL property of the scattered field  is just a consequence of the optical theorems of energy and helicity, and  does not presupose in the particle neither chirality ,  $\alpha_{em}= - \alpha_{me}$,  nor duality, $\epsilon^{-1}\alpha_{e}=\mu \alpha_{m}$ \cite{nieto1}.  Although the combination of both  theorems imposes \cite{nieto1} that  the existence of one these two latter properties of the particle polarizabilities   implies the other.

\textbullet  \, {\it In the near field zone, the scattered wave} in the  basis $({\bm \epsilon}_{\perp},{\bm \epsilon}_{\parallel}, {\bf s})$ has
\be
{\bf E}_{nf}({\bf r})= \frac{1}{\epsilon r^3} [3{\bf s}({\bf s}\cdot  {\bf p})-{\bf p}]  \,\,\,\,\, \, 
=-\frac{p_{\pm}}{\epsilon r^3} e^{\pm i\phi}[\pm i {\bm \epsilon}_{\perp}+cos  \theta \, {\bm \epsilon}_{\parallel} -2 sin \theta \,  {\bf s}]. \,\,\,\,\, \, \nonumber \\
{\bf B}_{nf}({\bf r})= \frac{\mu}{ r^3} [3{\bf s}({\bf s}\cdot  {\bf m})-{\bf m}] 
 =-\frac{\mu m_{\pm}}{r^3} e^{\pm i\phi}[\pm i {\bm \epsilon}_{\perp}+cos  \theta \, {\bm \epsilon}_{\parallel} -2sin \theta \,  {\bf s}] . \label{dipnf}
\ee

Thus this field  {\it being CPL}  at points ${\bf r}$ along the polar axis $OZ$, ( $\theta=0$,  or $\theta=\pi$).

\section{Excitation of helicity and energy  with general optical fields: The role of angular spectra with right circular and left circular polarization }
Returning to Eqs. (\ref{elipt1})-(\ref{pym_angul}) for general  optical fields, we have from the optical theorem for the helicity (\ref{todip4}) the following expression for its extinction from the incident field on scattering by the particle  induced dipole:
\be
\Im\{[{\bf p}_{+}({\bf r})+in {\bf m}_{+}({\bf r})] \cdot {\bf E}_{i}^{+\,*}({\bf r})-[{\bf p}_{-}({\bf r})-in {\bf m}_{-}({\bf r})] \cdot {\bf E}_{i}^{-\,*}({\bf r}) \}
+2\Re\{\alpha_{e} -n^{2}\alpha_{m}\}\nonumber \\
\times \Im\{ {\bf E}_{i}^{-}({\bf r}) \cdot {\bf E}_{i}^{+\,*}({\bf r})\}
=\frac{4k^3 n}{3\epsilon}\Im\{[{\bf p}_{+}({\bf r}) \cdot  {\bf m}_{+}^{ *}({\bf r}) + {\bf p}_{-}({\bf r}) \cdot  {\bf m}_{-}^{ *}({\bf r}) \} +CD({\bf r})+\frac{n}{2\pi c}{\cal W}_{\hel}^{a}. \,\,\,\,\,\,\,\,\,\, \label{topheli1}
\ee
While the extinction of incident energy is according to the standard optical theorem  (\ref{top}):
\be
\Im\{[{\bf p}_{+}({\bf r})+in {\bf m}_{+}({\bf r})] \cdot {\bf E}_{i}^{+\,*}({\bf r})+[{\bf p}_{-}({\bf r})-in {\bf m}_{-}({\bf r})] \cdot {\bf E}_{i}^{-\,*}({\bf r}) \} 
+2\Im( \alpha_{e} -n^{2}\alpha_{m} ) \,\,\,\,\,\,\,\,\,\,\,\,\,\,\,  \nonumber \\
\times\Re \{ {\bf E}_{i}^{-}({\bf r}) \cdot {\bf E}_{i}^{+\,*}({\bf r})\} 
=\frac{2k^3 }{3\epsilon}\{|{\bf p}_{+}({\bf r})|^2+ |{\bf p}_{-}({\bf r})|^2 +n^2[
|{\bf m}_{+}({\bf r})|^2+ |{\bf m}_{-}({\bf r})|^2 ]\}+CE({\bf r})+\frac{2}{\omega}{\cal W}^{a}. \,\, \label{topene1}
\ee
The terms $CD({\bf r})$ and $CE({\bf r})$ are
\be
CD({\bf r})=\frac{8k^3 n}{3\epsilon}[\Im\{(\alpha_{e} - n^2  \alpha_{m})\alpha_{me}^{*}\} \Re \{ {\bf E}_{i}^{-}({\bf r}) \cdot {\bf E}_{i}^{+\,*}({\bf r}) \}
- n \Im\{\alpha_{e}\alpha_{m}^{*}\}
 \Im\{ {\bf E}_{i}^{-}({\bf r}) \cdot {\bf E}_{i}^{+\,*}({\bf r})\}].  \,\,\,  \label{cd}
\ee
\be
CE({\bf r})=\frac{4k^3 }{3\epsilon}\{[|\alpha_{e}|^2 -n^4 |\alpha_{m}|^2)]
\Re \{ {\bf E}_{i}^{-}({\bf r}) \cdot {\bf E}_{i}^{+\,*}({\bf r})\} \,\,\,\,\,\,\, \nonumber \\
+2n \Re[ (\alpha_{e} -n^{2}\alpha_{m})\alpha_{me}^{*} ]
 \Im [ {\bf E}_{i}^{-}({\bf r}) \cdot {\bf E}_{i}^{+\,*}({\bf r})]\}.  \,\,\,\,\,\,\,\,\,\,\,\,\,\,\,   \label{ce}
\ee
In these equations ${\bf r}$ denotes the position vector of the  center of the particle inmersed in the illuminating field.  Now, in contrast with the scattering of an incident  elliptically polarized plane wave discussed above, the scattered helicity and energy convey interference between $ {\bf E}_{i}^{-}$ and  $ {\bf E}_{i}^{+}$.

Notice that by virtue of the asymptotic expression (\ref{asympt}), in the far-zone $CD(r\hat{\bf s})=CE(r \hat{\bf s})=0$ since ${\bm \epsilon}^{\pm\,*}(\hat{\bm s})\cdot {\bm \epsilon}^{\mp}(\hat{\bm s})=0$. It is also  interesting to observe from (\ref{topheli1})-
(\ref{ce}) that if the particle is dual, $\alpha_e=n^2 \alpha_m$,  the terms of interference between ${\bf  E}^{+}$ and  ${\bf  E}^{-}$ are zero and so are $CD({\bf r})$ and $CE({\bf r})$  for any ${\bf r}$. Then   (\ref{topheli1}) and  (\ref{topene1}) reduce to  equations similar to (\ref{topheli}) and  (\ref{topene}). 

However the important point is that now the appearence of  the interference factor  $[ {\bf E}_{i}^{-}({\bf r}) \cdot {\bf E}_{i}^{+\,*}({\bf r})]$ in  (\ref{topheli1}) and  (\ref{topene1}) allows one to choose the incident field such that   either  $\Im[ {\bf E}_{i}^{-} \cdot {\bf E}_{i}^{+\,*}]$ or  $\Re[ {\bf E}_{i}^{-} \cdot {\bf E}_{i}^{+\,*}]$ is zero, or small,  for  the helicity extinction (\ref{topheli1}) and for  the   intensity extinction (\ref{topene1}), respectively.  We thus shall analyse  the consecuences of $2\Re\{\alpha_{e} -n^{2}\alpha_{m}\} \Im\{ {\bf E}_{i}^{-} \cdot {\bf E}_{i}^{+\,*}\}$ or $2\Im( \alpha_{e} -n^{2}\alpha_{m} )\Re \{ {\bf E}_{i}^{-} \cdot {\bf E}_{i}^{+\,*}\}$  being non-zero  in  the left sides of (\ref{topheli1}) and  (\ref{topene1}), respectively,   as a consequence of the choice of illumination on the particle.

Using (\ref{constia}) the left sides of (\ref{topheli1}) and (\ref{topene1}) are in terms of the polarizabilities and fields:
\be
\Im\{[{\bf p}_{+}({\bf r})+in {\bf m}_{+}({\bf r})] \cdot {\bf E}_{i}^{+\,*}({\bf r})\pm[{\bf p}_{-}({\bf r})-in {\bf m}_{-}({\bf r})] \cdot {\bf E}_{i}^{-\,*}({\bf r}) \}= \,\,\,\,\,\,\,\,\,\, \nonumber \\
\Im\{\alpha_{e}+n^2 \alpha_{m}\} (|{\bf E}_{i}^{+}({\bf r})|^2 \mp |{\bf E}_{i}^{-}({\bf r})|^2) 
+2n\Re\{\alpha_{me}\}  (|{\bf E}_{i}^{+}({\bf r})|^2 \pm |{\bf E}_{i}^{-}({\bf r})|^2). \,\,\,\,\,\,\,\,\,\,\,\,\,\,\,\,  \label{lefttops}
\ee

Using (\ref{lefttops})  we now address  the rate of extinction of helicity   ${\cal W}_{\hel}^{s}$ (cf. Eq. (17) in \cite{nieto1})  and  energy  $ {\cal W}^{s}$  in the particle,  given by the left sides of (\ref{topheli1}) and (\ref{topene1}),    as  functions of the polarizabilities:
\be
\frac{\mu}{2\pi c} {\cal W}_{\hel}^{s} \equiv \Im\{[{\bf p}_{+}({\bf r})+in {\bf m}_{+}({\bf r})] \cdot {\bf E}_{i}^{+\,*}({\bf r}) 
-[{\bf p}_{-}({\bf r})-in {\bf m}_{-}({\bf r})] \cdot {\bf E}_{i}^{-\,*}({\bf r}) \} \,\,\,\,\,\,\, \nonumber \\
+2\Re\{\alpha_{e} -n^{2}\alpha_{m}\} \Im\{ {\bf E}_{i}^{-}({\bf r}) \cdot {\bf E}_{i}^{+\,*}({\bf r})\}
=\{\Im\{\alpha_{e}+n^2 \alpha_{m}\} (|{\bf E}_{i}^{+}({\bf r})|^2 -|{\bf E}_{i}^{-}({\bf r})|^2)\,\,\,\,\,\,\,\,\,\,\,\,\,\,\,\,\,\,\nonumber \\
+2n\Re\{\alpha_{me}\}  (|{\bf E}_{i}^{+}({\bf r})|^2 + |{\bf E}_{i}^{-}({\bf r})|^2)
+2\Re\{\alpha_{e} -n^{2}\alpha_{m}\}
 \Im\{ {\bf E}_{i}^{-}({\bf r}) \cdot {\bf E}_{i}^{+\,*}({\bf r})\}.\,\,\,\,\,\,\, \label{g}
\ee
And 
\be
\frac{2}{\omega} {\cal W}^{s} \equiv \Im\{[{\bf p}_{+}({\bf r})  
+in {\bf m}_{+}({\bf r})] \cdot {\bf E}_{i}^{+\,*}({\bf r})
+[{\bf p}_{-}({\bf r})-in {\bf m}_{-}({\bf r})] \cdot {\bf E}_{i}^{-\,*}({\bf r}) \}  \,\,\,\,\,\,\, \nonumber \\
+2\Im( \alpha_{e} -n^{2}\alpha_{m} )\Re \{ {\bf E}_{i}^{-}({\bf r}) \cdot {\bf E}_{i}^{+\,*}({\bf r})\}]
=  \{  \Im\{\alpha_{e}
+n^2 \alpha_{m}\} (|{\bf E}_{i}^{+}({\bf r})|^2 +|{\bf E}_{i}^{-}({\bf r})|^2) \,\,\,\,\,\,\,\,\,\,\,\,\nonumber \\
+2n\Re\{\alpha_{me}\}  (|{\bf E}_{i}^{+}({\bf r})|^2 - |{\bf E}_{i}^{-}({\bf r})|^2)
+2\Im\{\alpha_{e} -n^{2}\alpha_{m}\}
 \Re\{ {\bf E}_{i}^{-}({\bf r}) \cdot {\bf E}_{i}^{+\,*}({\bf r})\}  \}
. \,\,\,\,\,\,\,\,\,\,\, \label{g1a}
\ee
Notice that ${\cal W}_{\hel}^{s}\neq 0$ even if $\alpha_{me}=0$ and  $|{\bf E}_{i}^{+}({\bf r})|^2 =|{\bf E}_{i}^{-}({\bf r})|^2$. It should be remarked that  in the  particular case of  incident CPL  plane  waves, or CPL  beams without longitudinal component,   one has (choosing propagation along e.g. $OZ$):   ${\bf E}_{i}^{+}= {E}_{i}^{+}{\bm \epsilon}^{+}$,  ${\bf E}_{i}^{-}= {E}_{i}^{-}{\bm \epsilon}^{-}$; and  since $\Re \{ {\bf E}_{i}^{-} \cdot {\bf E}_{i}^{+\,*}\}= \Im \{ {\bf E}_{i}^{-} \cdot {\bf E}_{i}^{+\,*}\}=0$, Eq. (\ref{topheli1})  becomes (\ref{topheli})    and    Eq. (\ref{topene1})  reduces to (\ref{topene}). Hence, in this case  ${\bf E}_{i}^{+}$ and ${\bf E}_{i}^{-}$ do not interfere, and when  $|{\bf E}_{i}^{+}|=|{\bf E}_{i}^{-}|= |{\bf E}_{i}|$ Eqs. (\ref{g})  and  (\ref{g1a}) are similar to those of standard circular dichroism which our formulation shows that yields the rate of helicity extinction, first with an incident LCP wave, and then with one being RCP; both of the same amplitude. In such a situation (\ref{g}) and (\ref{g1a}) become respectively proportional to the well-known  numerator: $4n \alpha_{me}^{R}|{\bf E}_{i}({\bf r})|^2$ and denominator: $2(\alpha_{e}^{I} +n^2  \alpha_{m}^{I})|{\bf E}_{i}({\bf r})|^2$ of the CD {\it  dissymmetry factor} \cite{tang1,choi}. (The superscripts $R$ and $I$ denoting  real and imaginary part). 

However, our general equations  (\ref{g})  and  (\ref{g1a})  cover many other configurations, (in particular    those so-called superchiral fields \cite{tang1}, which is known, however,  to be limited to molecules with  $\alpha_{m}\simeq 0$ \cite{choi}). We next show the broader scope of  (\ref{g})  and  (\ref{g1a}) with chiral optical beams possessing a longitudinal component,  which as we shall show, plays a key role. We will see that according to whether one chooses such  illuminating beams yielding either $\Re\{ {\bf E}_{i}^{-} \cdot {\bf E}_{i}^{+\,*}\}=0 $ or $\Im\{ {\bf E}_{i}^{-} \cdot {\bf E}_{i}^{+\,*}\} = 0$ one respectively enhances the  extinction rate of helicity (\ref{g}) versus that of energy (\ref{g1a}), (and thus the ratio between them), or viceversa. Notice that since out of resonance the real part of the polarizabilities are usually greater than the imaginary parts, the last term of (\ref{g}) may be larger than that of  (\ref{g1a}). Hence, one may produce bigger enhancement in $ {\cal W}_{\hel}^{s}$ than in $ {\cal W}^{s}$ with those choices of  $\Re$ and $\Im$ of  $ {\bf E}_{i}^{-} \cdot {\bf E}_{i}^{+\,*}$.

\section{Optical beams whose angular spectrum representation contains left circular and right circular plane waves}
In the paraxial approximation $\partial_z \simeq ik_z$, so that the equation ${\bm \nabla} \cdot {\bf E}=0$ implies that $E_z=(i/k) {\bm \nabla}_{\perp} \cdot {\bf E}_{\perp}$ \cite{berry}; ($\perp$ denotes transversal, i.e. $XY$ component). The electric vector of an optical beam is then written in terms of its angular spectrum as \cite{mandel, nietolib}
\be
{\bf E}({\bf r})=e^{ikz}\int_{-\infty}^{\infty} {\bf e}({\bf s}_{\perp}) e^{ik{\bf s}_{\perp} \cdot {\bf R}} e^{-ik z\frac{|{\bf s}_{\perp}|^{2}}{2}}  d^2 {\bf s}_{\perp}. \label{beam1}
\ee
Having denoted  ${\bf r}=({\bf R}, z)$, ${\bf R}=(x,y,0)$, $ {\bf s}=({\bf s}_{\perp}, s_z)$, ${\bf s}_{\perp}=(s_x, s_y, 0)$. 

We shall consider  the Gaussian beam; i.e. the one from which other  fields, like Hermite and Laguerre-Gaussian beams, are generated \cite{zauderer}.

We  write for (\ref{beam1}) the decomposition (\ref{eheli}) of each   component into  LCP and RCP waves by expressing the Gaussian angular spectrum \cite{mandel,nietolib} as  \\
 $ {\bf e}({\bf s}_{\perp})= (k^2 W_{0}^{2}/4\pi)\exp [-k^2W_{0}^{2}|{\bf s}_{\perp}|^{2}/4][e_{0}^{+} {\bm \epsilon}^{+}({\bf s}) 
+e_{0}^{-} {\bm \epsilon}^{-}({\bf s})]$. \\ $e_{0}^{+}$ and $e_{0}^{-}$ being complex constants, and $W_0$ standing for the beam waist at $z=0$. Then we express the beam as:
\be
{\bf E}({\bf r})=\frac{(kW_{0})^{2}}{4\pi}e^{ikz}\int_{-\infty}^{\infty} e^{[-k^2 W_{0}^{2}|{\bf s}_{\perp}|^{2}/4]} e^{ik{\bf s}_{\perp} \cdot {\bf R}} e^{-ikz\frac{|{\bf s}_{\perp}|^{2} }{2}}
[e_{0}^{+} {\bm \epsilon}^{+}({\bf s}) +e_{0}^{-} {\bm \epsilon}^{-}({\bf s})]
 d^{2} {\bf s}_{\perp}.  \label{gauss1}
\ee
 Recalling that ${\bm \epsilon}^{\pm}({\bf s})=(1/\sqrt{2})(\hat{\bm \epsilon}_{\perp}({\bf s}), \pm i \hat{\bm \epsilon}_{\parallel}({\bf s}), 0)$, and writing in the Cartesian basis $\hat{\bm x},\hat{\bm y},\hat{\bm z}$, (see Fig. 1):  $\hat{\bm \epsilon}_{\perp}({\bf s})=(\sin \phi, \cos \phi, 0)$, \,\,\,\,\,
$\hat{\bm \epsilon}_{\parallel}({\bf s})=(\cos \theta \cos \phi, \cos \theta \sin \phi, -\sin \theta)$, $\\
{\bf s}=(\sin \theta \cos \phi, \sin \theta \sin \phi, \cos \theta)$,\,\,\,\,
$ d^{2} {\bf s}_{\perp}=d\Omega=\sin\theta \cos\theta d\theta d\phi$.\,\, $0\leq \theta\leq \pi$,  \,\, $0\leq \phi\leq 2\pi$

Performing the $\phi$ and $\theta$ integrals we obtain (see integrals 3.937.2 and 6.631.1 of \cite{gradshteyn}) after making  $\cos \theta \simeq 1$ in all factors of the integrand but not in the exponentials as involved in the paraxial approximation, and writing $x+i y=R \exp(i \Phi)$,  $\Phi$ being the azimuthal angle, we derive:
\be
{\bf E}({\bf r})=\frac{ W_{0}^{2}}{4 \sqrt{\pi}\, \sigma^3}e^{ikz} \{\frac{R}{2}\,
_{1}F_{1}(\frac{3}{2}; 2; -\frac{R^2}{2 \sigma^2}) [-e_{0}^{+}\exp(-i\Phi) (\hat{\bm x}+i\hat{\bm y}) 
+e_{0}^{-}\exp(i\Phi) 
 (\hat{\bm x}-i\hat{\bm y})] \,\,\,\,\,\,\,\,\nonumber   \\
+\frac{i}{k}  ( e_{0}^{+}-e_{0}^{-}) _1 F_{1}(\frac{3}{2}; 1; -\frac{R^2}{2 \sigma^2}) \hat{\bm z}\} . \,\,\,\,\,\,\,\,\,\,\,\,\,\,\,\,\,\,\,\,\label{gauss2}
\ee
$ _1 F_{1}$  is Kummer's confluent hypergeometric function \cite{abramowitz}. $\sigma^2=W_{0}^{2}/2 +iz/k$. Eq. (\ref{gauss2}) represents a hypergeometric beam which, containing LCP and RCP plane waves, differs from some previously put forward \cite{kotlyar}.   A generalization of this beam to  arbitrary index $m$ with vortices $\exp (\pm im\Phi)$ and topological charge $m$ is made by including a  factor $\exp(-im\phi)$  in $ {\bf e}({\bf s}_{\perp})$. Notice from  (\ref{gauss2}) that due to the paraxial approximation the transversal 
$XY$-component of  ${\bf E}$  is the sum of two fields  [cf. Eq. (\ref{ang masmen})]: one, ${\bf E}^{+}$, is LCP and has a complex amplitude proportional to  $-e_{0}^{+}$; the other, ${\bf E}^{-}$, is RCP and its amplitude factors $e_{0}^{-}$.  These two CPL beams also have a longitudinal component $E_z$, proportional to  $(i/k) e_{0}^{+}$ and    $-(i/k)e_{0}^{-}$, respectively,  as shown by the last term of  (\ref{gauss2}).   Next we see the relevance of this longitudinal component to control the dipole emission, enhancing the rate of either  helicity or energy extinction. Using (\ref{gauss2})  we obtain for the incident energy and helicity factors in the left side of (\ref{g})  and (\ref{g1a}) [we now  drop the subindex $i$ in those equations, understanding that the incident electric field is (\ref{gauss2})]:
\be
|{\bf E}^{+}({\bf r})|^2 \pm |{\bf E}^{-}({\bf r})|^2=\frac{W_{0}^{4}}{16\,\pi \sigma^6} (|e_{0}^{+}|^2 \pm |e_{0}^{-}|^2) [\frac{R^2}{2}\,  _{1}F_{1}^{2}(\frac{3}{2}; 2; -\frac{R^2}{2 \sigma^2})+\frac{1}{k^2} 
\,  _{1}F_{1}^{2}(\frac{3}{2}; 1; -\frac{R^2}{2 \sigma^2})] . \,\,\, \, \,\,\, \,\,\, \, \,\, \,\label{sumdifgauss}
\ee
Of course the choice of the upper or lower  sign in $\pm$ of (\ref {sumdifgauss})   yields the beam energy or the helicity [cf. Eq. (\ref{ang2})], respectively.
\begin{figure}[htbp]
\centerline{\includegraphics[width=0.6\columnwidth]{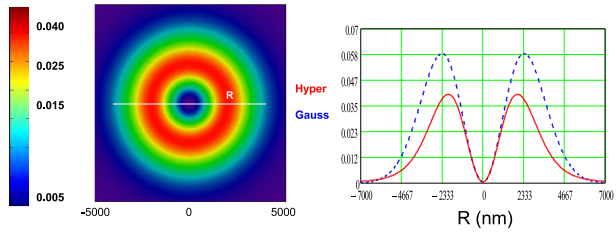}}
\caption{(Color online). Intensity   $|{\bf E}^{+}|^2 + |{\bf E}^{-}|^2$,  [cf. Eq.  (\ref{sumdifgauss})], at $z=0$ of the hypergeometric beam of  Eq. (\ref{gauss2}) . Left: Color map of the transversal distribution.  Right: a cut of this spatial  distribution as a function of the  coordinate $R$  along a diameter,  (full red).  The  distribution     when  the $_{1}F_{1}$ functions are replaced by a Gaussian of the same $\sigma$, is also shown, (broken blue line).}
\end{figure}

Fig. 2 shows the transversal intensity distribution  $|{\bf E}^{+}|^2 + |{\bf E}^{-}|^2$  of this beam, given by Eq. (\ref{sumdifgauss}) at $z=0$,  for $e_{0}^{-}=a e_{0}^{+}  \exp(i b \pi/2) $, $b$ real, $e_{0}^{+}=1$ (in arbitrary units) $a=1$ , $\lambda=589$ nm, $ W_{0}=4 \lambda$. This choice of  the value of   $ e_{0}^{+}$  and the presence of the factor $W_{0}^{2}/4 \sqrt{\pi}\, \sigma^3$ of the beam amplitude in (\ref {gauss1}) produces  small values of these intensities. Also since   $R^2>>\lambda^2$, apart from points  close to $R=0$  the second term of  (\ref{sumdifgauss}), given by the longitudinal component of the beam, hardly contributes to this intensity distribution. However as seen next, this longitudinal component becomes crucial when the   helicity, extinguished from the incident beam and  thus radiated or scattered -or absorved or converted \cite{norris}- by the particle,  is considered. For comparison, we also show  this intensity distribution when the $ _{1}F_{1}$ functions of (\ref{sumdifgauss}) are substituted by a Gaussian with the same value of  $\sigma^2$. The difference between both distributions is small due to the similar shapes of the Gaussian and these hypergeometric  functions.
 
On the other hand, the real (and imaginary)  part of the product ${\bf E}^{-}\cdot {\bf E}^{+}$ reduces to:
\be
\left \{ \begin{matrix}  \Re \\  \Im \\     \end{matrix}   \right \}
 [{\bf E}^{-}\cdot {\bf E}^{+\, *}]=\left \{ \begin{matrix}  \Re \\  \Im \\     \end{matrix}   \right \}[{ E}_{z}^{-}\cdot { E}_{z}^{+\,*}]
=-\frac{W_{0}^{4}}{16\,\pi \sigma^6 k^2}\, \left \{ \begin{matrix}  \Re \\  \Im \\     \end{matrix}   \right \} [e_{0}^{-}e_{0}^{+ \, *}]
\, _1 F_{1}^{2}(\frac{3}{2}; 1; -\frac{R^2}{2 \sigma^2})\} . \,\,\, \, \,\,\, \,\,\, \, \,\, \,\label{sumdif}
\ee
So that either of these quantities, $\Re[\cdot]$ or $\Im[\cdot]$, may  be made arbitrarily small (or zero) depending on the choice of parameters  $e_{0}^{-}$ and $e_{0}^{+}$  for the beam, which may make arbitrarily small (or zero) the factor
$\left \{ \begin{matrix}  \Re \\  \Im \\     \end{matrix}   \right \}  [e_{0}^{-}e_{0}^{+ \, *}]$.  In  the next section we show the relevance of this  choice in connection with Eqs.(\ref{g}) and (\ref{g1a}). For example, choosing like for Fig. 2:    $e_{0}^{-}/e_{0}^{+} = \pm a  \exp(i b \pi/2) $, $a$ and $b$ being  real, the value of $\left \{ \begin{matrix}  \Re \\  \Im \\     \end{matrix}   \right \} [{\bf E}^{-}\cdot {\bf E}^{+\, *}]$ will oscillate as \, $\mp \left \{ \begin{matrix}  \cos (b \pi/2) \\  \sin (b \pi/2) \\     \end{matrix}\right \}$,  thus possesing several zero values   in the interval $0\leq b \leq 4$.

Notice that a kind of Hermite and Laguerre Gaussian beam modes $(m,n)$ are straightforwardly worked out  from (\ref{gauss2}) on making upon ${\bf E}({\bf r})$ the operations: $\partial_{x}^{m} \partial_{y}^{n}$ and 
$(\partial_{x}+ i \partial_{y})^{m}(\partial_{x}- i \partial_{y})^{m+n}$, respectively \cite{zauderer}. Likewise,  Bessel beams with LCP and RCP angular components may be described by Eq. (\ref{gauss1}) using an angular spectrum : $ \delta ( {\bf s} - {\bf s}_0) [e_{0}^{+}(\hat {\bm \epsilon}_{\perp}({\bf s})+  i \hat {\bm \epsilon}_{\parallel}({\bf s})) 
+e_{0}^{-}(\hat {\bm \epsilon}_{\perp}({\bf s}) -  i \hat {\bm \epsilon}_{\parallel}({\bf s}))]$.
\section{ Example: Enhancing the emission of either  chirality or  energy}
As an illustration of the relevance of Eqs. (\ref{g}) and  (\ref{g1a}), we consider a helical molecule with $\alpha_{e}^{R}=1.04 \times 10^{-2}$ nm$^3$, $\alpha_{me}^{I}=6.2 \times 10^{-5}$ nm$^3$, $\alpha_{me}^{R}=0$,  in an environment with $\epsilon=\mu=1$  at an illumination wavelength $\lambda=589$ nm. $\alpha_{e}^{I}=(2/3) (2\pi/\lambda)^3(\alpha_{e}^{R})^2\simeq 0.96 \cdot 10^{-10}$nm$^3 \,<<\alpha_{e}^{R}$,   $|\alpha_{m}|< 10^{-5}|\alpha_{e}|$   \cite{barron1, capasso}.  

These polarizabilities yield according to (\ref{g}) and (\ref{g1a}) for the  helicity extinction  ${\cal W}_{\hel}^{s}$:
\be
\frac{\mu}{2\pi c} {\cal W}_{\hel}^{s}\simeq (\alpha_{e}^{I}+ \alpha_{m}^{I}) (|{\bf E}_{i}^{+}|^2 -|{\bf E}_{i}^{-}|^2)+2(\alpha_{e}^{R}-\alpha_{m}^{R})  \Im [ {\bf E}_{i}^{-} \cdot {\bf E}_{i}^{+\,*}];  \label{ggh}
\ee
and for the rate of energy extinction ${\cal W}^{s}$:
\be
\frac{2}{\omega} {\cal W}^{s}\simeq (\alpha_{e}^{I}+\alpha_{m}^{I}) (|{\bf E}_{i}^{+}|^2 +|{\bf E}_{i}^{-}|^2)+2(\alpha_{e}^{I} -\alpha_{m}^{I})
 \Re[ {\bf E}_{i}^{-} \cdot {\bf E}_{i}^{+\,*}]. \,\,\,\,\,\,\label{g1}
 \ee
\begin{figure}[htbp]
\centerline{\includegraphics[width=0.6\columnwidth]{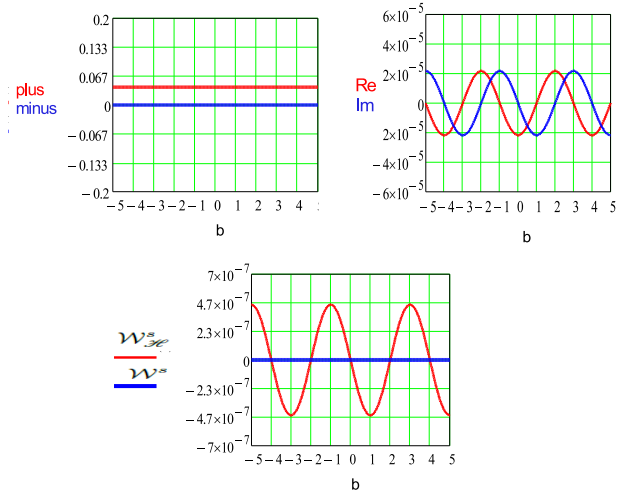}}
\caption{(Color online). Upper left: Intensity: $plus$ $=|{\bf E}^{+}|^2 + |{\bf E}^{-}|^2$ (red line) and helicity: $minus$ $=|{\bf E}^{+}|^2 - |{\bf E}^{-}|^2$ (blue line), [cf. Eq. (\ref{sumdifgauss})], as a function of $b$ for  the hypergeometric beam of  Eq. (\ref{gauss2}) at $z=0$ and  near the peak at: $R=2000$ nm, (see Fig. 2, right).  $e_{0}^{-}=a e_{0}^{+}   \exp(i b \pi/2) $, $e_{0}^{+}=1$ (in arbitrary units),  $a=1$.   \,\,  Upper right: $Re$ $=\Re  [{\bf E}^{-}\cdot {\bf E}^{+\, *}]$ and $Im$ $=  \Im [{\bf E}^{-}\cdot {\bf E}^{+\, *}]$, [cf Eq. (\ref{sumdif})], for the same beam and choice of parameters. Lower middle: Rate of helicity extinction ${\cal W}_{\hel}^{s}$ (full red)  and energy extinction ${\cal W}^{s}$ (broken blue) in terms of $b$.}
\end{figure}
We see from (\ref{ggh})  that objects with such a purely imaginary $\alpha_{me}$ would produce no signal  in a standard circular dichroism configuration, i.e. under illumination   with  plane CPL waves, for  which: $|{\bf E}_{i}^{+}|^2 =|{\bf E}_{i}^{-}|^2$, ${\bf E}^{-}\cdot {\bf E}^{+\, *}=0$.  (We
recall that in such experiments the objects  (molecules)   usually have  $\alpha_{me}^{R} \leq 10^{-3}  \alpha_{e}^{I}$, but $\alpha_{me}^{R} \neq 0)$.  However,  impinging the particle by LCP and RCP beams with  longitudinal component, like those of Eq. (\ref{gauss2}) , and for example choosing as above : $e_{0}^{-}/e_{0}^{+} = \pm a  \exp(i b \pi/2) $, Fig. 3 shows, at $R=2000$ nm and  $z=0$,  $|{\bf E}_{i}^{+}|^2 \pm |{\bf E}_{i}^{-}|^2$, as well as  $\Re \,$ (and $\Im$) of $[{\bf E}^{-}\cdot {\bf E}^{+\, *}]$    as functions of $b$   for $W_0=4\lambda$, $a=1$.  The incident  helicity, given by the quantity   $minus$ of Fig. 3, is zero since  $|{\bf E}_{i}^{+}|^2 =|{\bf E}_{i}^{-}|^2$. As seen, the oscillations of  the term $2(\alpha_{e}^{R}-\alpha_{m}^{R})  \Im [ {\bf E}_{i}^{-} \cdot {\bf E}_{i}^{+\,*}] $ of  (\ref{ggh}) and of $2(\alpha_{e}^{I} -\alpha_{m}^{I})
 \Re[ {\bf E}_{i}^{-} \cdot {\bf E}_{i}^{+\,*}]$  of (\ref{g1}) lead   to those of the helicity $ {\cal W}_{\hel}^{s}$  and energy $ {\cal W}^{s}$ extinction rate, respectively. The latter is constantly zero due to the very small value of the factor $(\alpha_{e}^{I} -\alpha_{m}^{I})$ for these polarizabilities. 

 The corresponding quotient between $ {\cal W}_{\hel}^{s}$  and $ {\cal W}^{s}$ would ve very large in this case. Therefore this is just an illustration of how such a ratio may be enhanced depending on the constitutive parameters of the particle and choice of the beam. 
Other objects with different values of the polarizabilities may yields similar enhancements of either the extinguished helicity -chirality- or  energy depending on whether  $\Im [{\bf E}^{-}\cdot {\bf E}^{+\, *}]$ dominates upon  $\Re [{\bf E}^{-}\cdot {\bf E}^{+\, *}]$ in (\ref{g})  and  (\ref{g1a}), or viceversa. 
For instance, were the "particle" magnetodielectric with  $\alpha_{m}^{I}$ comparable to $\alpha_{e}^{R}$,  or  just one or two orders of magnitude smaller, (a difficult case to deal with conventional circular dichroism \cite{choi}),  the factor $2(\alpha_{e}^{I} -\alpha_{m}^{I}) \Re[ {\bf E}_{i}^{-} \cdot {\bf E}_{i}^{+\,*}]$ will give rise to an  amplitude of the oscillations in  $ {\cal W}^{s}$ comparable to that in $ {\cal W}_{\hel}^{s}$, or one or two orders of magnitude lower. However, the phase shift of the oscillations of  $\Re[ {\bf E}_{i}^{-} \cdot {\bf E}_{i}^{+\,*}]$ and  $\Im[ {\bf E}_{i}^{-} \cdot {\bf E}_{i}^{+\,*}]$, (cf. $Re$ and $Im$ in Fig. 3), allows us to tailor the beam,  producing an enhancement of  $ {\cal W}_{\hel}^{s}$  or of $ {\cal W}^{s}$.

\section{Concluding remarks}
Based on a recent optical theorem put forward for the  electromagnetic helicity -or chirality- extinction  rate  in quasimonochromatic wavefields \cite{nieto1}, we have demonstrated  that dichroism is not only a manifestation of molecular absorption, but it  is a universal phenomenon which appears in the scattering -or  diffraction- of twisted waves. This provides a general basic answer to the question on the conditions under which an object produces chiral fields and/or dichroism, and whether a chiral  scatterer is required to produce such effect.

In this respect we have established that both  dichroism  and chirality of emitted or scattered  wavefields  from wide sense dipolar particles   are  consequences of the helicity of the illumination, or of the mutual relationship between the  emitting electric and magnetic dipoles;  but these phenomena do not require the object  constitutive parameters, refrative indices and polarizabilities,  to be those of a chiral structure. For example, as we have shown,  to obtain a circularly polarized emitted or scattered field,  it is a sufficient condition that the particle induced electric and magnetic dipoles rotate and differ from each other by only a $\pm \pi/2$ phase; but no chiral cross-polarizability $\alpha_{me}$ is necessary. Thus an achiral  particle ($\alpha_{me}=0$) may produce dichroism on scattering of a chiral incident wave. Henceforth  the standard concept of circular dichroism is generalized to include fields with both LCP and RCP components and a net helicity.

Based on the angular spectrum representation we have introduced new families of optical beams with  right circular and left circular polarization, and with longitudinal components.  Tailoring these fields, used  in our optical theorem as incident waves on the scattering particle, overcomes previous limitations of circular dichroism without  needing to place nearby  additional objects to enhance the signal \cite{klimov, dionne, kivshar}.  Depending on the parameters chosen for these beams, the enhancement of  the extinction rate of helicity and/or of  energy is produced,   i.e. the dichroism scattered signal is either augmented or lowered.  This not only provides a new  procedure for object (and particularly enantiomeric) characterization on illumination with twisted beams, but it also yields a way of controling  the helicity and energy   of radiated wavefields by  using such scattering particles as secondary sources.

\dataccess{This research involves no data.}

\funding{Work  supported by MINECO, grants FIS2012-36113-C03-03,  FIS2014-55563-REDC and FIS2015-69295-C3-1-P.}

\ack{ The author thanks Dr. J. M. Au\~{n}\'{o}n for a critical reading of the manuscript and helpful comments}



\end{document}